\documentclass{sig-alternate}

\usepackage{verbatim} 

\usepackage[lined,boxed,commentsnumbered,linesnumbered]{algorithm2e}

\usepackage{color}

\newcommand{\textred}[1]{\textcolor{red}{#1}}
\ifx\noeditingmarks\undefined
   \newcommand{\pgwrapper}[2]{\textred{#1 #2}}
\else
   \newcommand{\pgwrapper}[2]{}
\fi

\makeatletter
\let\@copyrightspace\relax
\makeatother

\begin{document}

\title{Elastic Pathing: Your Speed is Enough to Track You}

\numberofauthors{1}
\author{
\alignauthor
Bernhard Firner, Shridatt Sugrim, Yulong Yang, Janne Lindqvist\\
       \affaddr{Rutgers University}\\
       \affaddr{671 Route 1 South}\\
       \affaddr{North Brunswick, New Jersey, USA}\\
       \email{\{bfirner,ssugrim,yulong,janne\}@winlab.rutgers.edu}
       \email{Corresponding author: janne@winlab.rutgers.edu}
}

\maketitle
\begin{abstract}
Today people increasingly have the opportunity to opt-in to ``usage-based'' automotive insurance programs for reducing insurance premiums. In these programs, participants install devices in their vehicles that monitor their driving behavior, which raises some privacy concerns. Some devices collect fine-grained speed data to monitor driving habits.
Companies that use these devices claim that their approach is privacy-preserving because speedometer measurements do not have physical locations. However, we show that with knowledge of the user's home location, as the insurance companies have, speed data is sufficient to discover driving routes and destinations when trip data is collected over a period of weeks. To demonstrate the real-world applicability of our approach we applied our algorithm, {\em elastic pathing}, to data collected over hundreds of driving trips occurring over several months.  With this data and our approach, we were able to predict trip destinations to within 250 meters of ground truth in 10\% of the traces and within 500 meters in 20\% of the traces.  This result, combined with the amount of speed data that is being collected by insurance companies, constitutes a substantial breach of privacy because a person's regular driving pattern can be deduced with repeated examples of the same paths with just a few weeks of monitoring.
\end{abstract}




\section{Introduction}

In the past, insurance companies did not have a large amount of information about their customers so
everyone would pay similar prices, despite the large variations in driving habits.
Recently, technological advances have created ways to observe human driving behavior.
Taking advantage of this technology, some US-based insurance
companies~\cite{allstatedrivewise,gmacinsurance,onboardadvisor,progressivesnapshot,statefarmindrive}
now offer consumers vehicle insurance policies with the option of installing a tracking
device into their vehicle. This allows the insurance companies to charge less to safe
drivers, in an attempt to attract those people as customers. However, unlike a truck driver who is being monitored
while doing a job, these consumers will be monitored during all of their daily activities, public and
private.  Although some of these monitoring devices are based upon GPS information and offer no
privacy protections, (such as OnStar~\cite{OnStar}), many other devices and insurance programs are
advertised as being privacy-preserving. For instance, as an alternative
to GPS data some of these devices log speed information instead. In this paper,
we demonstrate that logging this timestamped speed data is also not privacy-preserving,
despite the insurance companies claims.

The privacy problem introduced in this paper is significant because the data collection by insurance companies is an ``always on'' activity. Data may be logged forever,
thus, deducing location data from speed traces represents a huge breach of privacy for drivers using
these programs. Even if insurance companies are not currently obtaining location traces from their
data today, this does not guarantee it will not present problems in the future. Because the data is not considered location data it is probably not being treated as such, and
may eventually be obtained by any number of antagonists that do know how to process the data
traces to obtain location information. For example, law enforcement agencies
may request the information as they have done with other data sets, e.g. electronic toll
records~\cite{tollrecords}. 

Breaching user privacy with just a starting point and speed information is a
difficult task, otherwise insurance companies would not claim that the
information being gathered is privacy preserving. Indeed, matching a single
speed trace to all of the roads in a country or state is extremely
difficult, but having the home address of a person, as the insurance
companies do, provides a starting location for some paths simplifying the problem.

\begin{figure}
\includegraphics[width=\linewidth]{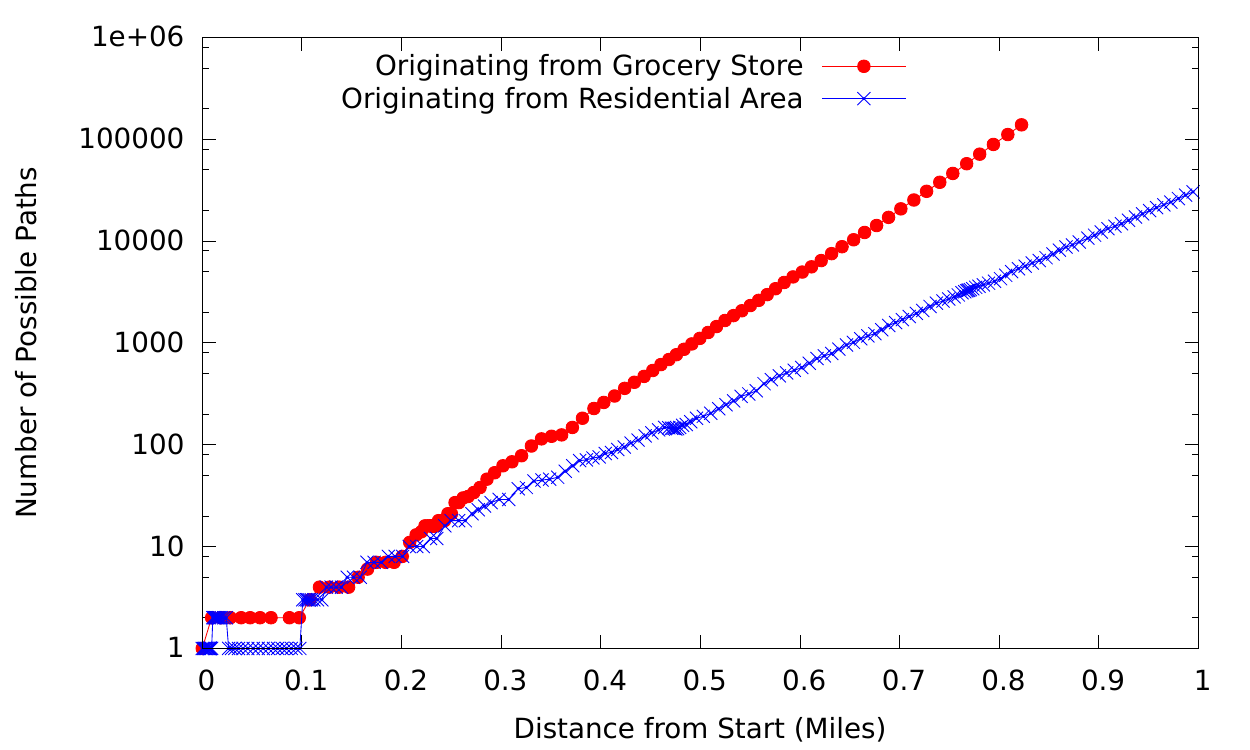}
\caption{\label{fig:path_growth} Number of path choices within a given distance of a starting location.
The pattern of growth on a log-linear scale shows that the number of paths increases exponentially,
making exploration of all possible paths infeasible. Within one mile there are over $10^6$ paths diverging from a grocery store
and over $10^4$ paths from a residential area. Locations central to major transportation routes,
like a grocery store, will see a higher increase in paths compared to a residential location.
Residential roads are also more likely to dead-end: in this example the number of paths from the
residence actually falls at the beginning because only one path does not dead-end immediately.}
\vspace{-6pt}
\end{figure}

Reproducing an exact driving path from just speed data and a starting location is challenging.
Speed data does not indicate if a person is turning
at an intersection or merely stopping at a stop sign or red light, so multiple
alternative pathways exist that match some of the speed data. Of course
multiple routes can be explored, but even within just a few minute's drive of a
person's home in our data set there may be thousands of unique paths that the person could have taken,
so exploring every single possible path is infeasible. The growth of
possible paths within a distance of just one mile of a starting location is illustrated
by Figure~\ref{fig:path_growth}. There may only be a few tens or hundreds of roads within a mile of a starting location, but there are many unique ways to travel along those roads. At one mile, there are over 100,000 possible paths
the driver could have taken when the trip starts from a grocery store next to several highways.

In addition to the explosion of possible paths the driver could have taken, the problem is made
hard by the fact that speed data is ambiguous; that is to say it does not uniquely match just a single path. For example, the speed of a vehicle making a right turn can also
match the speed of a vehicle making a left, or the speed of a vehicle that decelerates because of a
pot hole. There is no certain way to tell which way a car goes at an intersection with just speed
data, meaning that a segment of speed data may match multiple paths.

Since there are so many possible paths and speed data is ambiguous we must score paths according to some error metric
and seek out the final path with the least error. However, finding the optimal path from an
exponentially growing search space is not easy.  Simple solutions, such as greedily
following a single path, may end in failure without even finding any solution; for instance when the path chosen by
a greedy algorithm suddenly dead ends.  Later in the paper we discuss several other approaches that do
not work.

This paper makes two major contributions.  
\begin{enumerate}
\item We present a novel algorithm -- \emph{elastic pathing} -- 
to extract location traces from speedometer data and starting locations.
\item Applying this algorithm and to real-world traces, we show how the data collected by many insurance companies is not
privacy preserving despite their claims. 
\end{enumerate}

\emph{ To best of our knowledge,
our work is the first to demonstrate how people's locations can be discovered based only on timestamped
speed data and knowledge of a single starting location.}
We will also show that even without predicting path
endpoints with 100\% accuracy long-term data collection allows an antagonist to eventually identify
physical regions that are frequently visited. The combination of our algorithm with analysis of
long-term driving traces can obtain private information from what is currently considered privacy-preserving data.  
As minor contributions, we have collected real-world traces over several months that
serve as a testing set for our approach, we offer suggestions for future work that may improve
path prediction further, and provide a discussion of privacy-preserving alternatives to the
collection of speedometer data that do not require cryptography to implement.

\section{Privacy Issues with Usage-Based \\ Automotive Insurance}
\label{sec:usage}
In several countries, including countries in North America, Europe and Japan, some insurance
companies \cite{Troncoso:2011:PPP:2015838.2015879} have introduced ``usage-based'' insurance
programs. In addition to basing insurance premiums on the characteristics of the car and history and
knowledge about the drivers, usage-based insurance programs also take into account how and when the
car is driven. The incentive to participate by insurance policy holders is a potential discount to
their premiums for good driving behavior.  In addition to providing potential premium discounts,
some companies also provide their customers statistical data, such as overviews of driving habits
and information about how much they drive per day or per week.  In the past most of these programs
used to track drivers' locations with GPS \cite{Troncoso:2011:PPP:2015838.2015879}, but currently
there are several programs that claim to be privacy preserving because they use other kinds of data.

In this paper, we focus on a more recent type of a program, which is enabled by collecting
speedometer among other data directly from the car. The idea is that by not collecting GPS data and
instead collecting other driving information, such as speedometer data, these approaches are more
privacy-preserving since the insurance companies are not collecting drivers' locations. 
There are several insurance companies in the US that use this method.  For example, the
Snapshot device provided by Progressive and the DriveWise device provided by Allstate will record
vehicle speed data while not collecting GPS locations \cite{allstatedrivewise,progressivesnapshot}.

The data in these programs  is collected by connecting a device (provided by the insurance company)
to the car diagnostic port OBD-II~\cite{USEPA}.  Since 1996, OBD-II has been made mandatory for all
cars sold in the United States~\cite{USEPA}. The European Union has also made the European
equivalent EOBD~\cite{eurodirective} mandatory for both petrol and diesel engined cars in Europe.
This standard interface allows for collecting comprehensive amounts of diagnostic information.

The data that is collected varies with each insurance company, but in the cases discussed above, it
includes time and speedometer data.  The devices will regularly upload this data to the insurance
company's servers via a mobile data link. The driving habits that some insurance companies state
that they are interested in include ``How often you make hard brakes, how many miles you drive each
day and how often you drive between midnight and 4 a.m.'' There is also a very important subtlety here. 
For example, Progressive claims they do not collect the data about when people are speeding
(which some people in the public interpret as meaning that speed data is not collected at all).
Instead, they \emph{do} collect speed data, but it follows from the logic that since they do not
know your location, they do not know if are you speeding or not.

For our analysis of the privacy concerns of these programs we need to know the
sampling rate of the data collection devices, but the companies do not disclose on their web sites
how often do they sample e.g. speedometer data. However, e.g.  Allstate \cite{allstatedrivewise}
includes the following information ``hard braking events are recorded when your vehicle decelerates
more than 8 mph in one second (11.7 $ft/s^2$) [and] extreme braking events are recorded when your
vehicle decelerates more than 10 mph in one second (14.6 $ft/s^2$).''
We also know that companies will sample data much faster than they need to.  In the United
States federal vehicular safety regulations~\cite{USmotorsafety393} mandate that vehicles traveling
at 20 mph must be able to brake to a full stop in 20 feet at a deceleration rate of 21 feet/second.
Since we know what kind of events the companies are interested in and the maximum performance of
vehicles on the road, we can simply estimate the bounds on sampling rates based upon the amount of
information required to detect these events.

The events that are to be detected occur over a time interval of just one second
($\tau = 1$). Using the Nyquist sampling theorem, we know that we must sample at twice this rate to
detect these features, a rate of two samples per second.
  This is a lower bound on any feature detection, and companies may sample at significantly higher
rates to detect other features. Consider that a driver that tailgates very closely will often tap
their own brakes for very short time durations. If insurance companies wished to detect these events
they would need to sample at a rate that was twice the speed of these brief taps.

\subsection{Summary of Privacy Issues}

To summarize the privacy issues, the insurance companies using speedometer based measurements are claiming that they cannot find people's locations
based on their collected data. Our work shows this is not true. We will show in further sections that a lot private location information is indeed obtainable, contrary to the claim of 100\% privacy protection. Further, by finding out correct paths on the roads, the insurance companies could also deduce subtle issues such as were the drivers speeding,
despite their claims to the contrary.

We will show that our elastic pathing algorithm is able to do more than just identify the endpoint of a trip -- it also eliminates a
large number of locations where the trip could not have ended. This is also a serious privacy concern because
it allows to identify changes in regular routines very easily. For example, if a
person usually drives for groceries after work but the predicted path goes in a different direction, then it is
highly likely the driver is breaking their routine.

Indeed, previous research has shown that location (derived from GPS data) can be automatically analyzed
to produce profiles of driver's behavior, social activities and work activities \cite{iqbalworkshop}.
A comprehensive discussion on the the consequences of breaches to location privacy are given
by Blumberg and Eckersley \cite{efflocationalprivacy}. They discuss how location databases can
be used by others to ask questions such as ``Did you go to an anti-war rally on Tuesday?'', 
``Did you see an AIDS counselor?'', ``Have you been checking into a motel at lunchtimes?'',
``Why was your secretary with you?'', or ``Which church do you attend? Which mosque? Which gay bars?''

\section{Related Work}
\label{sec:related}

A lot of work has concentrated on anonymizing or obfuscating~\cite{Popa:2011:PAL:2046707.2046781,Gruteser:2003:AUL:1066116.1189037,Krumm:2009:RDT:1560004.1560010,Brush:2010:EEU:1864349.1864381,krumm:inference} location traces
because location traces contain a great deal of behavioral information that
people consider private~\cite{Brush:2010:EEU:1864349.1864381,ludford,patil,tsai}. Krumm~\cite{Krumm:2009:SCL:1569346.1569363} has written an overview of computational location privacy techniques, and Zang and Bolot~\cite{Zang:2011:ALD:2030613.2030630} have
recently questioned the possibility of releasing privacy-preserving cell phone records while still maintaining
research utility in those records.  Accordingly, predicting future mobility patterns and paths from human mobility
traces is a well-explored topic, including the following results.

Based on analysis of location data, such as mobile phone cell tower locations, researchers have shown
1) the nature of individual mobility patterns is bounded, and that humans visit only a few locations
most of the time (e.g., just two~\cite{barabasi:individual, Eagle_Pentland_2009,Golle:2009:AHL:1560004.1560038}); 2) that there
is high level of predictability for future and current locations
(e.g.~\cite{barabasi:limitspredictability, Farrahi:2011} -- most trivially, ``70\% of the time the
most visited location coincides with the user's actual
location''~\cite{barabasi:limitspredictability}); and 3) that mobility patterns can also be used to
predict social network ties (e.g,~\cite{Wang:2011:HMS:2020408.2020581}). In a more applied domain,
GPS traces have been used to learn transportation modes (e.g. walking, in a bus, or
driving)~\cite{Zheng:2010:UTM:1658373.1658374}, for predicting turns, route, and
destination~\cite{Ziebart:2008:NLC:1409635.1409678}, for predicting family routines (e.g. picking up
or dropping of children to activities)~\cite{Davidoff:2011:LPP:1978942.1979119}, probabilistic
models describing when people are home or away~\cite{krumm:learning}, and recommending friends and
locations~\cite{Zheng:2011:RFL:1921591.1921596}.  However, none of the above work can be used to discover
locations based solely on speedometer measurements and a user's starting location.

Related work in the field of dead reckoning does suggest that speedometer traces should have some
level of information that can be extracted. Dead reckoning works by using speed or movement data and
a known location to deduce a movement path.
Dead reckoning has typically been used for map building with mobile
robots~\cite{elastic_correction} or as an addition to
Global Navigation Satellite Systems (GNSS) such as GPS~\cite{kalmann_integration,zhao2003extended}.
When supplementing GNSS data, odometer data for dead reckoning might only be used when GNSS
information is unavailable, such as when a vehicle passes through a tunnel or an area with many tall
buildings. However, this kind of dead reckoning cannot work without frequent location ground truths
as there is no perfect method to match speed data to turns in a road.

Map building with robots is interesting because there are often no exact ground truths, only
location estimates. Golfarelli et al.~\cite{elastic_correction} describe a method that can assemble
maps from arbitrary distance estimates to and from landmarks. The problem of
deducing a traveled path from only speed data and a starting point is, in some ways, the reverse of
this map building problem: we already have the map, but have difficulty identifying landmarks (in
this case turns) from just speed data.

Ubiquitous computing devices help people during their daily lives in different
contexts, for instance by monitoring our activity and fitness levels or interesting events social
that we have attended.  However, the ubiquitous nature of these monitoring devices has security and
privacy ramifications that are not always considered. For
example, a lack of sufficient protection in the Nike+iPod sport kit allowed other people to
wirelessly monitor and track users during their daily
activities~\cite{Saponas:2007:DTY:1362903.1362908}.  Although these kinds of devices clearly appeal
to consumers, the potential loss of privacy is a growing concern.

In the specific area of usage-based insurance, the
privacy concerns of these insurance programs have been studied before by Troncoso et al.~\cite{Troncoso:2011:PPP:2015838.2015879}. 
However, this work dates back to schemes which would send raw location (GPS) coordinates to either insurance providers or brokers, and Troncoso et al. proposed
a cryptographic scheme PriPAYD~\cite{Troncoso:2011:PPP:2015838.2015879} to address the problem. Our work shows that speedometer-based solutions, which
were not considered by Troncoso et al., are not privacy-preserving either. 

Finally, most closely related to our work are side-channel attacks that use accelerometer
data from smartphones~\cite{Aviv:2012:PAS:2420950.2420957}. Projects such as
ACComplice~\cite{accomplice} and AutoWitness~\cite{Guha:2010:ALT:1869983.1869988} have used
accelerometers and gyroscopes for localization of drivers. However, the information from the
smartphone can be used to actually detect when turns occur. In contrast, we have only a time series
of speed data available. While this speed data might indicate when a vehicle stops or slows down at
an intersection, unlike accelerometer data, it does not indicate if any turn is taken.

\section{Data Collection}
\label{sec:collection}

To test the hypothesis that a driving route can be reconstructed from a known starting point and a
speed trace we needed to collect a large body of testing data. We require both a ground truth of GPS
data and a sample set of speed data.  This data can be collected with the approach that the
insurance companies use, by connecting a device to the Onboard Diagnostics standard OBD-II
connector.  Devices connected to the car this way will receive accurate information about the car's
speed directly from the vehicle's computer and can pair these speed values with GPS information to
provide an accurate ground truth.  To log this information we used a cell phone with GPS and a
Bluetooth enabled ODB-II diagnostic device.  We have also logged a much higher volume of GPS-only
data, from which we can reconstruct the speed trace.

Using the speedometer data from the ODB-II device is straightforward because raw speed data is
immediately available. However, collecting data using the ODB-II device required obtaining the
devices and was more cumbersome; in the rush of daily activities it was easier to use just GPS data
gathered from a smartphone. This data requires a small amount of processing to obtain speed values
from latitude/longitude pairs.  This was a two step process, as outlined below.

First, we used the haversine formula (see appendix) to approximate distances from the raw GPS coordinates.
Next we simply divide this value by the time interval (in fractions of an hour) to get an
instantaneous speed value for each time interval. Each speed value is then given a time stamp and
converted into the same format as the speedometer traces. We note that we did not see differences
in the accuracy of our algorithm between the two approaches of data collection.

\subsection{Description of Speed Data}
We had seven volunteers collecting data. Five volunteers collected GPS-only data over the period of several months, and two
volunteers gathered both GPS and speedometer data with an ODB-II device over a period of one month.  There
were 240 data traces that we used for pathing, totaling nearly 1200 miles (nearly 2000 km) of
driving data collected, with a median trip length of 4.65 miles (7.5 km), minimum trip distance of
0.38 miles (0.62 km), and maximum trip length of 9.96 miles (16.0 km).  These traces comprise 46
unique destinations, with more than half of these destinations visited multiple times by individual
drivers.

We believe this represents a diverse testing set; if our algorithm works well on this data, with all
of its variety (e.g. the differences between driver's habits), then it should perform similarly in
many other driving scenarios.  In addition to the differences between driving habits of
participating drivers, there are also differences in their vehicles. Examples of vehicles we used in
the study are small sedans, sports utility vehicles, and a pickup truck. We feel that this
represents a good sample of driving habits and vehicle types, so a pathing algorithm that works
properly on all of these traces will work correctly across the driving public as a whole.
Additionally, the insurance companies know the make and model of the vehicle they are insuring and
might be able to use additional information, such as the turning radius of particular vehicles, that
we do not use.

\subsection{Description of Map Data}
Map data was collected from the OpenStreetMap (OSM) project \cite{openstreetmap} using their open
REST API.  We parsed this data into node adjacency lists, turn restrictions, and other road
characteristics (e.g. speed, directionality) and stored the data in a local SQL database for rapid
querying.

The open street map data is divisible into two main types: \textit{nodes} and \textit{ways}.  Nodes
are numbered latitude/longitude pairs. They may document significant map features like bends in
roads or intersections. Nodes may consist of multiple attributes but they always include a latitude,
longitude and a unique node id. Ways are collections of node ids and some context tags that identify
features of the road that the group of nodes describe.  When ways cross the nodes at the
intersection belong to both ways, indicating that a turn is possible. Turn restrictions are
indicated by the way-ness of the road (one-way or two-way) and additional meta-information, such as
``no u-turn'' or ``only right turn''.

\section{Elastic Pathing}
\label{sec:methods}
In this section, we describe the algorithm we use to recreate a person's driving path given data trace with
speed/time tuples. Our approach relies upon two features of vehicular travel. First, there are
physical limitations to a vehicle's turning radius at high speeds, so a vehicle that travels a
particular path must travel at speed at or below the maximum speed possible.  Second, people will
only stop when they need to, e.g. at traffic lights and stop signs, until they reach their final
destination. This second assumption is not always true; a person may need to stop because of actions
of other vehicles, road construction, accidents, etc. However, when we continually observe a driver
only a small number of traces will have extraordinary events. 

We use data from Open Street Maps (OSM) to identify possible routes and locations of turns and we
find the route that best fits the given speed data.  OSM data is in the form of nodes that are
associated with different roads so a path will be a sequence of nodes. Each node has a physical
location, given by a latitude and longitude, allowing us to determine distances between segments of
road and the turning angle of a path of nodes.

Each sample from the data trace corresponds to some amount of distance traveled from one node to
the next along a path. The OSM data provides adjacency data for each node, so once a node is reached,
we must choose which node to begin moving towards. For instance, a four-way intersection will have a
single node at the intersection and four adjacent nodes.  Since we reach the intersection from one
direction we have three possible next nodes from which to choose.

Any pathing algorithm must follow some basic steps. First, as discussed previously and illustrated
in Figure~\ref{fig:path_growth}, the growth of paths is too great to explore all possible paths. Therefore,
any algorithm must have some methods of comparing paths and choosing ``better'' paths that more
closely match the given speed trace. This also means that the algorithm must keep a list of possible
paths, while also either limiting growth of the number of paths or removing paths as they become too
plentiful.

Another problem to overcome is that no path perfectly matches the speed data, because drivers
swerve around objects in the road, or take turns more widely or sharply than we expect.
Increasingly the estimation of a vehicle's progress along a path will become incorrect and mismatches
might make even the correct path seem impossible. For example, if the progress along a segment of
road is ahead of the vehicle's actual position then the vehicle might be going too quickly to make a
turn, but if we somehow corrected the distance traveled to account for some of these distance
estimation errors then we would find that the speed traces lines up perfectly with the turn.  Thus,
any algorithm must also correct paths as it explores them, trying to take into account these
variations in the travel distance.

One could think of many ways to attempt this pathing, and we thought of and explored several
approaches that do not work before finding an approach that gave surprisingly good results to
be harmful for people's privacy. Before describing our novel approach, we will first document the failed approaches to provide more background information to the community.

\subsection{Failed Approaches}
If deducing a driving route from speed data and a starting location was easy then insurance
companies would not claim that their data collection was privacy preserving. Although such path reconstruction
may seem simple in concept, it is actually very difficult. The first major difficulty is detecting turns with only speed data.
When someone breaks and then accelerates, did they make a turn or did they slow down because the
person in front of them was turning? If a person comes to a complete stop and then accelerates, did
they go straight or turn? When a person stops at an intersection are they the car closest to the
stop light, or are they behind ten other cars? The speed data alone does not give enough
information, and a single incorrectly identified turn will result in an error possibly as large as
the entire trip distance. Even if a turn is identified from the speed data, if the predicted
position of the car is off by a few meters then the wrong intersecting road may be chosen as the
destination of a turn.

In principle, turns could be predicted probabilistically based upon past driver
behavior~\cite{froehlich2008route}. However, we assume that we do not already know information about
an individual driver, so we cannot build a model of their specific driving. We could consider using a model
developed using many other drivers in the same area~\cite{krumm2010will}, but that has the
disadvantage of predicting the most common turns and paths rather than the unique path of an
individual. Estimating that a person probably shops at a supermarket near them or that they probably
work at the largest employer in their town is not necessarily a considerable breach of their privacy. Instead, we
identify specific locations that a particular individual visits, and are especially interested in
those places that are unique to the individual and not common to many people.

Since the heart of this pathing problem is the lack of turn information one approach is to treat
this as a classification problem. Here, we examine the speed traces to extract features: stop, right
turn, left turn, or straight road. If classification would work well, we can assign probabilities to
different choices at each intersection and eventually find the highest probability path.
However, road conditions and driving habits are very diverse. With our data set we were
unable to classify turn features with high enough accuracy to build correct paths. The sheer
volume of paths to explore meant that unless the correct path was extremely different from all of
the incorrect paths, the algorithm would find some incorrect paths that looked just as good, or better,
than the correct path.

Another approach we tried was to switch from a probabilistic model to a model where we identified time
windows where turns could occur and did not follow any paths that would make turns outside of these
windows. We expected this to reduce the number of paths we needed to explore, perhaps leaving just a
few paths at completion. However, because of the accumulation of distance estimation errors, turns in
the speed data do not align perfectly with the map data so we need to widen the time intervals
where turns were allowed. As the time windows widen the number of accepted paths
also increased, until we had too many possible paths to distinguish a single best fit.

We also tried a dynamic programming approach, keeping track of the score of the best possible
path at each node of the ways in the OSM data. However, there is a subtle flaw in this approach.
Whatever error metric is used for each path, that error will slowly increase as the path grows in
size due to the slight discrepancies in estimated and real travel distance. If there are two
possible paths to reach a node then this dynamic programming approach will tend to favor the shorter
path that reaches that node. However, if we explored the path further we might find that the longer
path with a worse score at the previous node better matches the road going forward -- but with this
dynamic programming approach we would have already dropped that path because it had the worse score.
This error would be unrecoverable.

The approach that works must keep track of the best path in terms of time; if two paths have the
same error score but one path has traveled through more of the speed samples then that path is
better than the other (up to that point) because it is accumulating less error per unit time.

\subsection{Our Approach: Elastic Pathing}
Next, we describe our successful approach we call \emph{elastic pathing}. This algorithm is based around
the observation that as we attempt to match the speed data to a path we stretch or compress the
predicted distance moved. However, after reconciling differences between a section of road and the
speed trace we must ``pin'' the path at that point (which we call a landmark) because any movement would
break apart the newly reconciled speed/time trace and distance data.
For instance, if the speed data goes to 0 indicating a stop where there is no intersection
we might pull the path forward by some distance to reach an intersection. After this has been done
the path from this point and earlier cannot be moved since it aligns with a feature in the road.
We call this approach \emph{elastic pathing} because of the stretching and compressing of the speed
trace to fit the road is conceptually similar to stretching a piece of elastic along a path while
pinning it into place at different points. To help understanding the algorithm, we introduce
a small set of definitions.

\IncMargin{1.5em}
\begin{algorithm}[t!]
  \SetAlgoLined
  \DontPrintSemicolon
  \SetKwData {Samples}{Samples}
  \SetKwData {StartNode}{StartNode}
  \SetKwData {PartialPaths}{Partial}
  \SetKwData {CompletePaths}{Complete}
  \SetKwData {NewPaths}{P'}
  \SetKwData {TypeA}{TypeA}
  \SetKwData {ErrorThreshold}{$\delta$}
  \SetKwData {CurPath}{CurPath}
  \SetKwFunction{sort}{sort}
  \SetKwFunction{join}{join}
  \SetKwFunction{dropLowerQuantile}{dropLowerQuantile}
  \SetKwFunction{gotoBranch}{gotoBranch}
  \KwIn{The starting point for pathing, \StartNode, and a set of speed samples, \Samples, and a
threshold within the best possible score to accept, \ErrorThreshold}
  \KwResult {\CompletePaths = list of possible paths within \ErrorThreshold of the best possible}
  \Begin {
    \PartialPaths $\longleftarrow \{[$\StartNode$]\}$\;
    \CompletePaths $\longleftarrow \emptyset$\;
    \While{\CompletePaths $= \emptyset$ OR \\\PartialPaths.first.error $< \ErrorThreshold \times$ \CompletePaths.first.error}{
      \NewPaths $\longleftarrow $ \gotoBranch(\PartialPaths.pop)\;
      \join(\CompletePaths, $\lbrace x \in \NewPaths \mid x$ complete $\rbrace$)\;
      \join(\PartialPaths, $\lbrace x \in \NewPaths \mid x$ incomplete $\rbrace$)\;
      \sort(\PartialPaths)
    }
  }
  \caption{\label{alg:pathing}Pseudo code for the elastic pathing algorithm.}
\end{algorithm}
\DecMargin{1.5em}

\begin{itemize}
\item[] \emph{Calculated Distance} The distance a vehicle traveled calculated from the speed and
time values in the speed trace.
\item[] \emph{Predicted Distance} The distance along a possible route on the road at a certain time
in the speed trace.
\item[] \emph{Error} The difference between calculated distance and predicted distance of a possible
route.
\item[] \emph{Feature} A vehicle stop in the speed trace or an intersection in the road.
\item[] \emph{Landmark} A place where the speed trace and road data match, but would become unmatched
if any stretching or compression of the predicted distance moved occurs. This includes the
previously mentioned feature, but also any other place where a mismatch between speed and road data
can occur.
\end{itemize}

\IncMargin{1.5em}
\begin{algorithm}[t!]
  \SetAlgoLined
  \DontPrintSemicolon
  \SetKwData {PP}{P}
  \SetKwData {ZZZ}{S}
  \SetKwData {ii}{i}
  \SetKwData {Back}{Back}
  \SetKwData {Fore}{Fore}
  \SetKwData {Edge}{Edge}
  \SetKwData {Branches}{P'}
  \SetKwFunction{Pin}{Pin}
  \SetKwFunction{advanceA}{compressA}
  \SetKwFunction{rewindA}{stretchA}
  \SetKwFunction{advanceB}{compressB}
  \SetKwFunction{rewindB}{stretchB}
  \SetKwFunction{join}{join}
  \SetKwFunction{intersection}{intersection}
  \SetKwFunction{maxSpeed}{maxSpeed}
  \KwIn{A path, \PP, speed samples, \ZZZ, and the current index into the speed samples, \ii}
  \While{\ii $<$ \ZZZ.length}{
    \tcp{Match turns at intersections to slower speeds.}
    \If{\PP at intersection}{
      \Branches $\longleftarrow \emptyset$\;
      \ForEach{$\Edge \in $\intersection(\PP)}{
        \If{$\ZZZ[\ii]$.speed $\leq$ \maxSpeed(\PP, \Edge)}{
          \Pin(\PP)\;
          \join(\Branches, $\PP\cup\Edge$)
        }
        \Else{
          \Fore $\longleftarrow$ \advanceA(\PP)\;
          \Back $\longleftarrow$ \rewindA(\PP)\;
          \join(\Branches, \Fore)\;
          \join(\Branches, \Back)\;
        }
      }
      \Return \Branches\;
    }
    \tcp{Match 0 speeds to intersections.}
    \If{$\ZZZ[\ii]$.speed $\approx 0$}{
      \While{$\ZZZ[\ii]$.speed $\approx 0$ AND\\\ii $<$ \ZZZ.length}{
        $++\ii$\;
      }
      \If{\PP at intersection}{
        \Pin(\PP)\;
        \Return \PP\;
      }
      \Else{
        \Fore $\longleftarrow$ \advanceB(\PP)\;
        \Back $\longleftarrow$ \rewindB(\PP)\;
        \Return \{\Fore, \Back\}\;
      }
    }
    $++\ii$\;
  }

  \caption{\label{alg:gotobranch}Pseudo code for the \emph{gotoBranch} function used in the elastic pathing
algorithm. This code advances a single path until it reaches discrepancy between the speed trace and
road. When this happens the path is corrected with compression or stretching and the path is
``pinned'' at what we call a landmark and the path's error is recomputed.  Multiple possible paths
may be returned at intersections.}
\end{algorithm}
\DecMargin{1.5em}

We can compute the fit of a path by the amount of stretching and compressing that needed to be done
to make the speed data match the path. The more stretching and compressing along a path the worse
its score.  At each iteration of the algorithm we sort all of the partial paths by their current
scores and then explore the path with the best score. That path advances until it reaches a feature
that requires a ``pin'' operation. At this point there may be multiple ways to advance the path so
several new paths may be created. Each new path's score is adjusted to reflect the stretching or
compression of the distance traveled from the last pinned landmark and the algorithm proceeds to
the next iteration.  The algorithm follows with the best score at each iteration, so the first path
to finish cannot be worse than any other path.  The pseudo-code for this algorithm is given in
Algorithm~\ref{alg:pathing}.

The most complicated operation done in the \emph{ElasticPathing} algorithm is \emph{gotoBranch}
function, whose pseudo code appears in Algorithm~\ref{alg:gotobranch}. The \emph{gotoBranch} function
advances a path until it comes to a feature on the map and returns every possible branching from
that feature. There are two features: an intersection in the road of the path and a zero speed
section of the speed samples.

When a path reaches an intersection it explores every possible direction. If the path is going at a
speed where it can move along the curve in the road then a landmark is set at that location with the
\emph{Pin} function. If the curve in the road is too great for the current speed then the distance
moved from the last landmark is compressed with the \emph{compressA} function and stretched with the
\emph{stretchA} function. Thus, when there is a mismatch between the speed samples and road data there
are two ways to resolve it and thus two new possible paths.

A similar situation occurs when the speed data indicates that the vehicle has come to a stop. If the
path is already at an intersection then the landmark is set with the \emph{Pin} function. Otherwise
the two solutions are found with the \emph{compressB} and \emph{stretchB} functions.  Rather than
compressing or stretching the speed trace as in the previous compress and stretch functions, these
functions compress or stretch the route data from the last landmark until an intersection
is found to match the zero speed segment.

After any landmarks are set, the ratio of stretching or compression from the last landmark increases
the error of each path. This means that a preexisting path may have less error than the current paths
and should be explored instead, so all possible branches are returned to the elastic pathing
function, the paths are placed in sorted order by their error, and pathing continues with the new
best path.

There are several details that we do not show in the pseudo-code. For instance, when we check if we
are at an intersection we allow space for the number of lanes in the intersecting road and
offset for another car in front of our vehicle. Determining the maximum speed of a turn is done by
assuming the maximum allowed incline in the road (7\%) \cite{roesstrafficengineering}, the turn radius of the road
based upon the number of lanes in the road and typical lane widths \cite{aasho}, and the
turn angle equation \cite{roesstrafficengineering}.

\section{Results}
\label{sec:results}

\begin{figure}[t!]
\centering
\includegraphics[width=\linewidth]{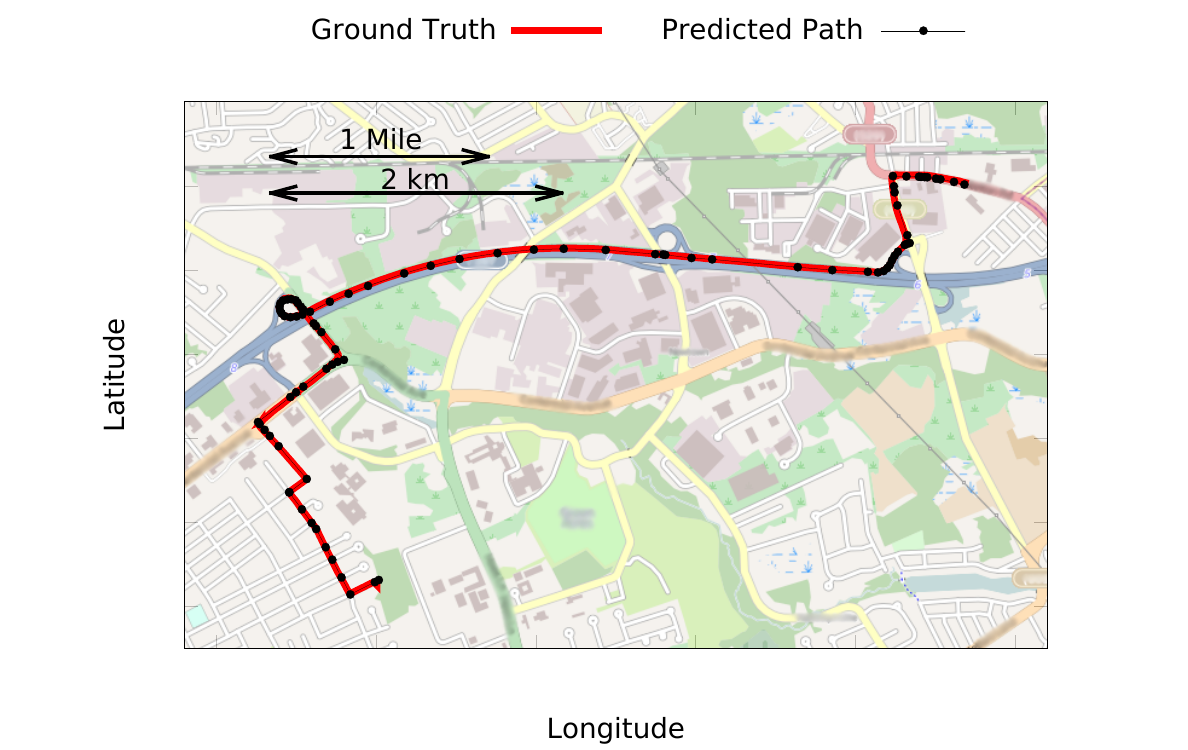}
\caption{\label{fig:best_case}The best possible case, a path from a shopping store to a home with no error. The
ground truth is indicated with a solid line while the nodes along the predicted are shown as dots.
In this case the predicted and actual paths match perfectly.}
\vspace{-6pt}
\end{figure}

We generated results for 240 traces comprising 46 unique destinations of 7 different individuals. Of
these destinations, 28 were visited more than once and 10 locations were
visited more than 5 times. The median trip length was 4.65 miles (7.5 km) with a
minimum trip distance of 0.38 miles (0.62 km) and maximum trip length of 9.96 miles (16.0 km).
The 2009 National Household Travel Survey NHTS found the average drive distance is 9.62 miles on weekdays and 10.03 miles on weekends~\cite{nhts_2009}, so our drivers had short, but not unreasonable commutes.
Traces were taken at many
different times of day and in varying road conditions over several months, and we believe that these
traces are representative of common driving behavior.

A Ruby implementation of the elastic-pathing algorithm processed all of the traces in under 30
minutes on a 2-core 2.2GHz machine, with an average running time of less than ten seconds per trace.
Even without an implementation in a high-performance language, the algorithm is able to process
traces far faster than they are generated. Since data analysis could essentially be done in
real-time as a vehicle's speed trace is being collected, the limiting factor on the time delay
between data recording and path prediction is probably the time overhead of collecting,
transmitting, and preprocessing (e.g. matching a trace to a specific driver and their known
locations) speed traces.
This means that it may be entirely feasible to do near real-time tracking of drivers with just speed
data, although the tracking will not be 100\% accurate.

\begin{figure}
\centering
\includegraphics[width=0.9\linewidth]{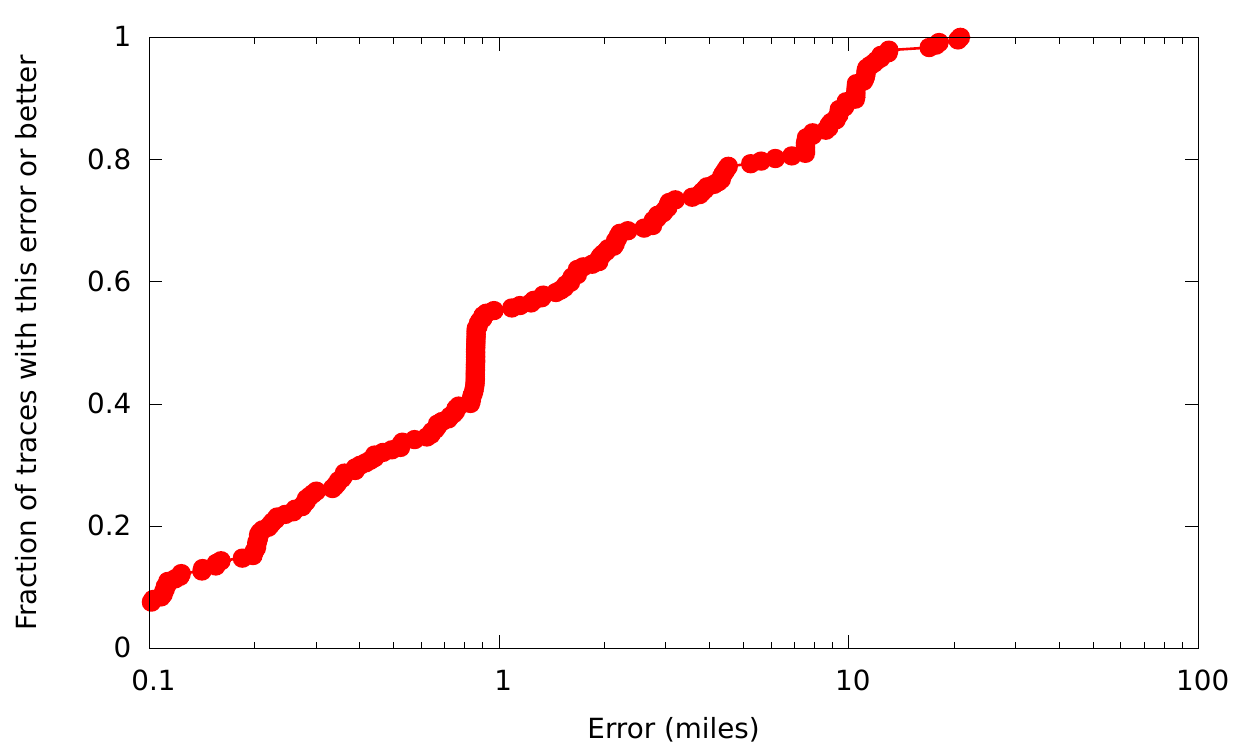}
\caption{\label{fig:cdf}The CDF of the distance from the pathing algorithm's endpoint compared to
the ground truth's endpoint. In 10\% of traces the algorithm found an endpoint less than 250 meters
from its actual location and 30\% of the traces were within 800 meters of the actual end point.
With thousands of driving traces gathered per individual per year this gives an insurance company a
very detailed view of an individual's lifestyle and habits.
Results did vary with the individual; the top 30\% of one driver's traces were all within 500 meters
while that same percentage were only within 1300 meters for another driver. 
}
\vspace{-6pt}
\end{figure}

The CDF of errors is shown in Figure~\ref{fig:cdf}.  In 10\% of the traces, the pathing algorithm
found an endpoint that was within 250 meters of the actual endpoint and 30\% of the traces were
within 800 meters of the actual endpoint. This distance is easily accurate enough to tell when a
person goes home, what areas a person drives through, and can identify many destinations of a
driver.  Since the size of a parking lot in a movie theater or shopping center is comparable to
these error values, this level of detail is enough to gain an in-depth view of an individual's
lifestyle and habits. When we consider that individuals often take multiple trips to the same
locations it is clear that, over time, every location that is commonly visited will be identified.

\begin{figure}
\centering
\includegraphics[width=0.9\linewidth]{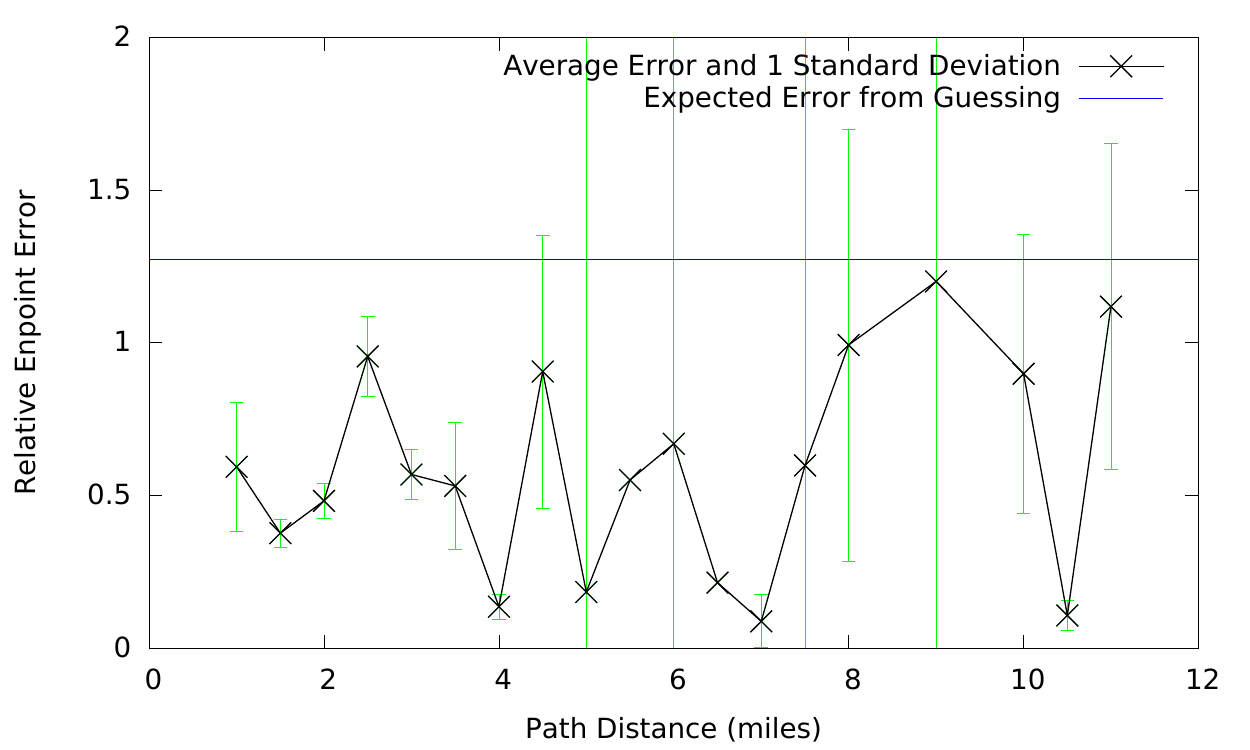}
\caption{\label{fig:relative_error} The average error of endpoints relative to trip distance.
This figure shows that there is no noticeable decrease in the
performance of the elastic pathing algorithm with trip distance. Manual analysis of our algorithm's
performance shows that prediction accuracy is more strongly
influenced by the route taken and traffic encountered than by the trip distance.
The expected error from guessing is simply the distance from the starting point to a random
location along the circumference of a circle centered at the starting point and with radius equal
to the trip distance.
}
\vspace{-6pt}
\end{figure}

The relative inaccuracy of our approach does not go up with trip distance, as shown in
Figure~\ref{fig:relative_error}. The percent of predicted endpoints within 250 meters of the actual
endpoint also does not decrease with distance in our data set, with trips as long as 10.5 miles
still having endpoints correctly predicted to within 250 meters. Our approach clearly does much better than
random guessing and almost always correctly identifies at least the general direction of travel.
The direction of travel is not a very serious privacy concern though, so we will use another filtering step to screen out
``noise'' from incorrectly predicted points.

\begin{figure*}[t]
\centering
\includegraphics[width=0.8\linewidth]{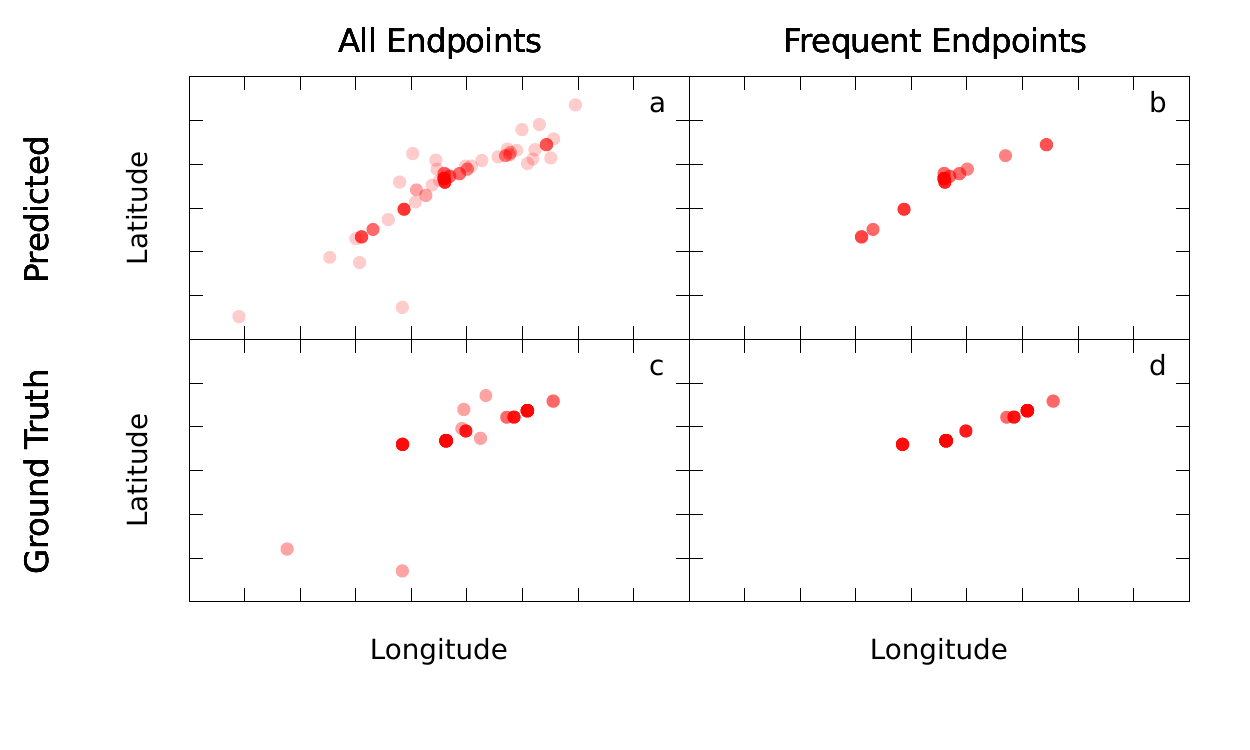}
\caption{\label{fig:endpoint_circles}These plots show how repeated trips to the same endpoint
reinforce that location. Circles (with radius 500m) represent endpoints for one individual's trips.
The darkness of each circle indicates the frequency of that endpoint. The top row of plots (a,b) shows
the predicted endpoints while the bottom row (c,d) shows ground truths. The first column (a,c) shows all
endpoints, while the second shows only those endpoints that were visited at least five times.
Although prediction error causes the set of all predicted endpoints to be noisy (a) reducing this
map to the frequently visited endpoints (b) creates a set of locations that closely resembles the
groundtruth map of frequent endpoints (d).  }
\vspace{-6pt}
\end{figure*}

We can further take advantage of multiple available traces for each driver to screen out points that
may be noise. We note that erroneous predictions are unlikely to repeatedly find the same wrong
location and instead usually make mistakes in different locations. Thus, we can use the frequency
that locations are predicted as a way to remove poor predictions and focus only on common
destinations, as shown in Figure~\ref{fig:endpoint_circles}.  The first row of the figure,
Figure~\ref{fig:endpoint_circles}(a,b), shows predicted endpoints while the second row,
Figure~\ref{fig:endpoint_circles}(c,d), shows ground truths. The first column,
Figure~\ref{fig:endpoint_circles}(a,c), shows all endpoints, while the second column,
Figure~\ref{fig:endpoint_circles}(b,d), shows only the endpoints that occurred with high frequency
in the traces. Although the initial endpoint predictions (Figure~\ref{fig:endpoint_circles}a) are
very noisy, filtering the endpoints by frequency in Figure~\ref{fig:endpoint_circles}b creates a
collection of endpoints that is fairly close to the ground truth in
Figure~\ref{fig:endpoint_circles}d.

We also noted that some individuals were more easily tracked than others. Although the top 10\% of
traces for traces from all individuals had a maximum error of 250 meters, for one individual the top
10\% had a maximum error error of only 164 meters, and the top 30\% of those individual's traces
still had errors of less than 500 meters. These individuals actually commuted along the same major
road and lived in similar areas, so the differing performance of the pathing algorithm on these two
individuals is likely to stem from differences in their personal driving characteristics and the
driving characteristics of their vehicles than from the areas where they drove. The practical
meaning of this is that learning the common locations and habits of some individuals might be much
easier than for others.

The pathing algorithm is able to do more than just identify the endpoint of a trip -- it also eliminates a
large number of locations where the trip could not have ended. This is also a serious privacy concern because
it allows an antagonist to identify changes in regular routines very easily. For example, if a
person usually goes home after work but the predicted path goes in a different direction then it is
highly likely the person is breaking their routine.

When the algorithm has enough information and a route is very unique -- for instance the spacing
between intersections on the road travelled does not match the spacing in any other road -- the
algorithm can do very well. This is illustrated in Figure~\ref{fig:best_case}. However, this case is
not normal, and at the very end when the algorithm chooses the correct left turn it may just as well
have chosen to turn right or gone straight. However, even if the algorithm made the wrong choice it
would still be close to 100 meters from the actual destination.

\begin{itemize}
	\item We believe the probability of error is closely tied to a few factors:
	\begin{itemize}
		\item Homogeneity of roads. If every road seems the same (same speed, similar
intersection intervals, etc), especially as in areas built in a grid, then they cannot be
distinguished.
    \item Traces with only slow speeds. Without high speeds to rule out turns and constrain
paths to a few major roads the correct path is indistinguishable from any other.
    \item Unpredictable stops from traffic, construction, etc.
	\end{itemize}
\end{itemize}

The largest barrier to improved performance is distinguishing one road from another.  If two roads
have similar features (e.g. same speed, similarly spaced intersections) then there is little way for
the algorithm to distinguish between the two. This is also affected by the traffic pattern of the
trip; if a vehicle stops at every intersection because of red lights then there is a great deal of
information about the spacing of those intersections. If, however, the vehicle stops at no lights
then there is no information about the spacing of intersection on the current road.  With more prior
information we may be able to do better -- for instance, if we knew the average waiting time at
different lights then the waiting time at an intersection might distinguish one road from another
even if a vehicle only stops at one or two lights.

This additional information, such as the average wait time at certain lights, or the average traffic
speeds of different roads during different times of day, may already be available to insurance
companies and this information is not difficult to gather.  If the algorithm had this information
available it could score different turning choices with higher accuracy, leading to improved
results.

\section{Discussion}
The implications of our results are that, given just timestamped speed traces and a starting
location, the privacy of a driver will be eroded as more of their speed traces are collected.
This is even though our approach does not yet predict traces and trip endpoints with 100\% accuracy. 
Driving habits are often regular and locations are visited multiple times, so an antagonist gets many chances to correctly
identify traveled paths. As a person's driving traces are collected and analyzed, a set of possible destinations will be built for that person.
A set of thousands of possible destination will be reduced to a handful. Incorrect endpoint
predictions are noisy and unlikely to repeat, so over time the endpoints that are predicted the most
are increasingly likely to be correct predictions of a person's common destinations.
Thus, we believe that the large body
of data already collected and still being collected would allow an antagonist to deduce many details
about a person's regular routine and irregular events. Starting from a person's home address an antagonist could identify
work locations, commonly visited stores, locations associated with regular hobbies, etc.
Additionally, any breaks in routines would be very obvious because the details that could be
obtained from a speed trace, such as distance traveled, duration of trip, and predicted endpoint of
the trip, would all change.

It is possible for companies to adjust their sampling rates to reduce
the potential for privacy breaches, however, there is no guarantee that someone could not improve upon our
algorithm and obtain results better than ours even with a lower sampling rate. Perhaps the best way
to improve the privacy protection of these systems is to do data processing in the car before
sending it to a remote server, so that only the end results are ever transmitted and stored.

\subsection{Privacy-Preserving Alternatives}
Although it is tempting for a company to capture as much information as
possible in case that data becomes useful later, this also presents
unnecessary privacy risks since.
In the case of auto insurance companies, a speed trace is not vital to this
system, as the real goal is the detection of unsafe driving habits, which can
be detected with a more privacy preserving set of driving features.

There are alternatives to collecting speedometer readings which should have much better privacy
protection.  In fact, some usage-based insurance programs are already offering more
privacy-preserving products.  Companies such as GMAC insurance, MileMeter and Corona Direct offer
programs that only require the mileage information from vehicles \cite{gmacinsurance, milemeter, coronadirect}.
PAYD coverage provided by Hollard Insurance Company collects mileage information and
driver style information~\cite{hollardinsurance}. Companies like PayGo Systems collect driving zone
information
instead of fine-grained location, besides basic information including
mileage and minutes of use~\cite{paygosystem}. 

The best practice for such monitoring systems is to only record relevant
data.  For example, devices could check speed data but only
send a notification when an even of interest was detected. In general, this
approach could be applied to other data collection methods, including GPS.

\begin{figure}[t]
\includegraphics[width=0.9\linewidth]{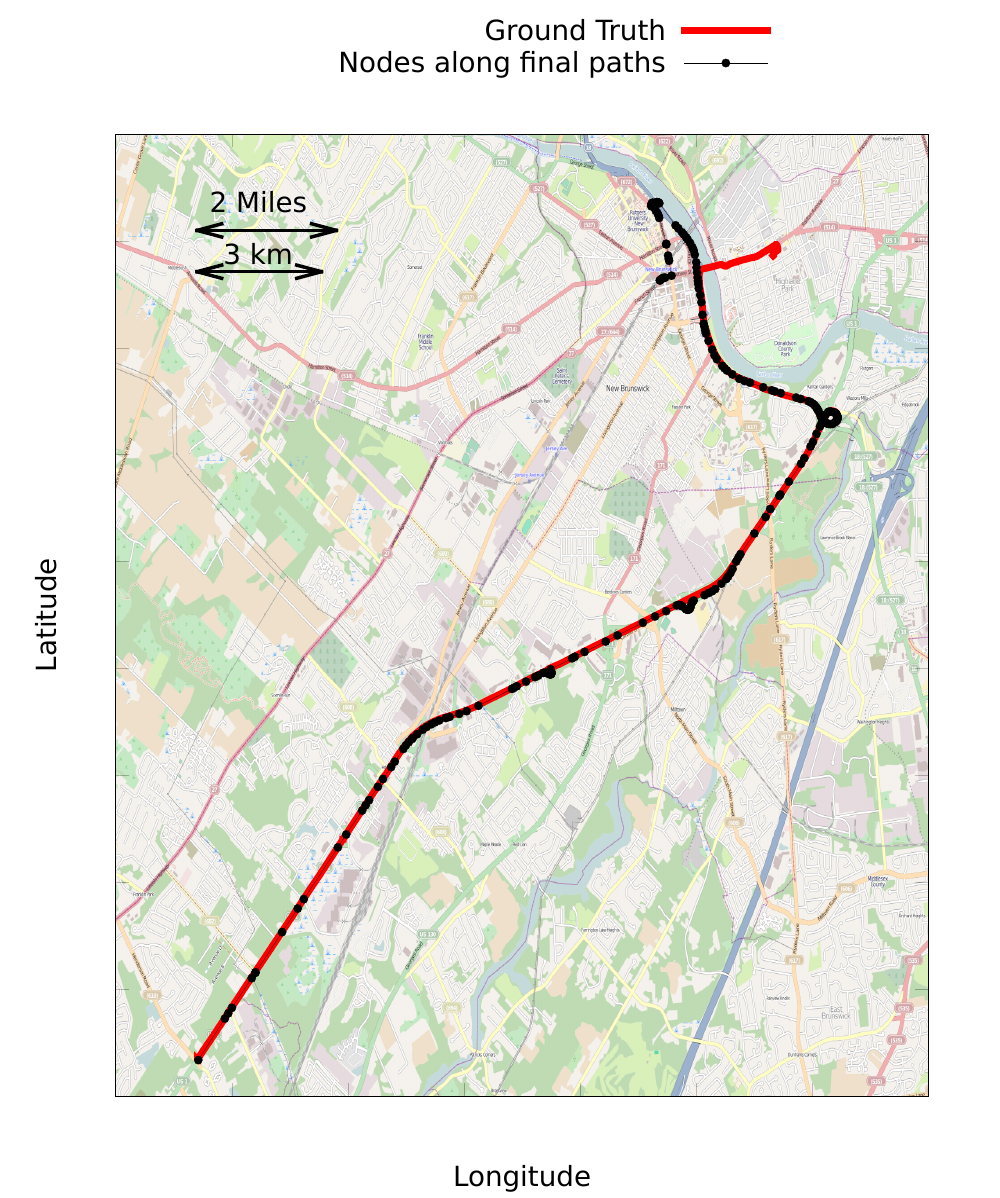}
\caption{\label{fig:follow_highway}An example path where the algorithm correctly follows
most of a path but predicts the wrong turn when the speed trace runs out of features.}
\vspace{-12pt}
\end{figure}

\subsection{Limitations and Improvements} Our main limitation is the difficulty distinguishing between similar roads. When a
vehicle stops at an intersection and then accelerates slowly a turn or straight
path are both possible, so our algorithm relies upon later features in the
speed trace to determine which of the possible routes was correct.
At the ends of paths there are simply too few features to make this
distinction.  At the beginning of paths this may simply cause there to be too
many possible paths to sort through.  These problems are exacerbated in
residential areas, where repeated, similar intersections are common.

We notice that travel through residential areas tends to occur at either ends
of paths rather than during the middle, because traveling through a residential
area is unlikely to be the fastest route to a destination.  Fixing mistakes in
residential areas by incorporating additional knowledge about the area may
greatly improve our algorithm's success since we had little difficulty
reconstructing paths along major roads, but tend to make mistakes once
the path leaves major roads. An example of this is in Figure~\ref{fig:follow_highway}.

There are two ways these errors could be addressed. First, the street addresses
of customers are available so the algorithm could check for paths to a person's
known home address at the end of predicted paths. Second, at the beginnings of
paths drivers move from local roads to larger arteries, so the algorithm could
also attempt to simply take the shortest paths to major roads if there are too
many ambiguous turn choices at the beginning of a path.

When accepted for publication, we will open source our work, which contains
tools to get the needed map data from Open Street Maps, and implementations of
our algorithms. We will also include
 a subset of the collected traces that has been explicitly authorized to be made publicly available on CRAWDAD or similar archival sites. 

\section{Conclusions}
We have demonstrated an attack against user location privacy that uses speed data from driving traces and
an initial starting location.  This problem seems simple in concept, but is
difficult in execution largely because turn detection with speed data is often ambiguous.  For
instance, if someone breaks and then accelerates they could have turned or they could have slowed
down because the person in front of them was turning. We have presented the design and implementation
of a novel elastic pathing algorithm and used a large corpus of real-world traces to show its accuracy.
Our trials showed that the algorithm is fast, and could also be used for real-time tracking of
people just based on their speed data.

Our results show that with a large corpus of data an entity can errode individuals' privacy
and gain information about typical routes and destinations.
In 20\% of individual traces we
predicted the correct trip end point to within 500 meters.
This means that a location visited
daily can be identified in about a week; locations visited on a weekly basis could be
identified with slightly more than a month of data.

Variations in daily routines are easier to identify than endpoints. Even when the algorithm does not
correctly find the proper end point it does correctly follow much of the actual travelled path. If
that predicted path breaks from a normal routine, such as going home after work, then a malicious
entity may be able to either deduce where the person could have gone with some foreknowledge of the
person's habits or take advantage of their altered routine.

In time, we expect that improvements will be made to our initial approach.
The data, once collected, does not go away and a continually improving path prediction algorithm may
plumb more private information from these traces. These traces should be treated as any
other private information instead of being handled as already anonymized data. Our work is not directed
against any company or organization -- experiences have shown that even well-intentioned uses of
data
can result in losses of privacy so we wish to highlight potential dangers of this type of data
collection.

Usage-based insurance has been introduced in at least 39 states of the United States, and more are
approving the approach because it has been claimed to be privacy-preserving. Our work shows that
speed-based data collection erode user privacy of policy holders over time.
We do not claim that insurance companies are violating policy holder privacy in this way.  However,
the general principle of any privacy preserving data collection is to collect only the data that is necessary for a
particular application, and no more.

\newpage

\section*{Acknowledgments}
This material is based upon work supported by the National Science Foundation under Grant Numbers 1228777 and 1211079. Any opinions, findings, and conclusions or recommendations expressed in this material are those of the author(s) and do not necessarily reflect the views of the National Science Foundation.

\appendix
\section{Haversine formula}
\begin{eqnarray}
A &=& (sin^{2}(\frac{\Delta lat}{2}) + sin^{2}(\frac{\Delta lon}{2}))  \times cos(lat_1) \nonumber \\
 & & \times cos(lat_2) \nonumber  \\
C &=& 2 \times atan2(\sqrt{A},\sqrt{1-A}) \nonumber \\
Distance &=& R \times C \times 0.62
\end{eqnarray}

Where $\Delta lat, \Delta lon$ are the changes in latitude and longitude respectively, and $R$ is the
radius of the earth (aproximately: 6,371km). The factor of 0.62 is for km to miles conversion.


\end{document}